\documentclass[9pt,twocolumn,twoside]{osajnl}
\usepackage{bm}
\journal{ol} 

\setboolean{shortarticle}{true} 

\title{Suspended silicon waveguides for long-wave infrared wavelengths}

\author[1,*]{J. Soler Penadés}
\author[2]{A. Sánchez-Postigo}
\author[1]{M. Nedeljkovic}
\author[2]{A. Ortega-Moñux}
\author[2]{J. G. Wangüemert-Pérez}
\author[1]{Y. Xu}
\author[2]{R. Halir}
\author[1]{Z. Qu}
\author[1]{A. Z. Khokhar}
\author[1]{A. Osman}
\author[1]{W. Cao}
\author[1,3]{C.G. Littlejohns}
\author[4]{P. Cheben}
\author[2]{I. Molina-Fernández}
\author[1]{G. Z. Mashanovich}

\affil[1]{Optoelectronics Research Centre, University of Southampton, Southampton SO17 1BJ, United Kingdom}
\affil[2]{Universidad de Málaga, Dept. de Ingeniería de Comunicaciones, ETSI Telecomunicación, Campus de Teatinos, 29071 Málaga, Spain}
\affil[3]{NOVITAS, School of Electrical and Electronic Engineering, Nanyang Technological University, Singapore 639798, Republic of Singapore}
\affil[4]{National Research Council Canada, Building M-50, Ottawa, K1A 0R6 Canada}

\affil[*]{Corresponding author: jsp1a15@soton.ac.uk}

\ociscodes{(130.0130) Integrated optics, infrared; (130.6010) Semiconductors; (050.5524) Subwavelength structures; (050.2770) Gratings.}

\begin{abstract}
In this paper we report suspended silicon waveguides operating at a wavelength of 7.67 \bm{$\mu$}m with a propagation loss of \bm{$3.1\pm0.3$} dB/cm. To our knowledge this is the first demonstration of low loss silicon waveguides at such a long wavelength, with loss comparable to other platforms that use more exotic materials. The suspended Si waveguide core is supported by a sub-wavelength grating that provides lateral optical confinement, while also allowing access to the buried oxide layer so that it can be wet etched using hydrofluoric acid. We also demonstrate low loss waveguide bends and s-bends.
\end{abstract}

\setboolean{displaycopyright}{true}

\begin{document}
	
\maketitle

Mid-infrared (MIR) photonics research has experienced rapid development in the last few years, which has been sparked by increased interest in the commercial, military and security applications of the field, and by much improved availability of characterization and testing equipment that is able to reach longer MIR wavelengths. Some of these applications, such as absorption spectroscopy of gases and chemicals in the fingerprint region, require the use of wavelengths above 7 $\mu$m \cite{Soref:06LWIR} (e.g. Sulfur dioxide has strong absorption near 7.6 $\mu$m \cite{Website:SO2NIST}), and there is therefore a need to develop photonic integrated circuits that can work throughout the MIR.

Silicon-on-insulator (SOI) is the dominant waveguide platform in silicon photonics, but in the MIR alternative platforms are needed because SiO$_2$ has extremely high absorption at wavelengths above 4 $\mu$m (i.e. higher than  2.5 $\times 10^3$ dB/cm at 7.6 $\mu$m, rising to $\sim$2.5 $\times 10^4$ dB/cm at 8.0 $\mu$m) \cite{Kitamura:07}. Several alternative MIR group-IV material waveguide platforms with wider transparency ranges have been reported in the last few years, such as silicon-on-sapphire \cite{Baehr-Jones:10,Liu:10,Li:11,Singh:15}, silicon-on-nitride \cite{Lin:13}, germanium-on-nitride \cite{Li:16}, graded Si/Ge \cite{Ramirez:17,Brun:14}, and germanium-on-silicon (GOS) \cite{Milos:15Ge,Malik:13,Chang:12}. So far group-IV material waveguiding has been demonstrated up to the 7.5-8.5 $\mu$m range, where graded Si/Ge \cite{Vakarin:17,Brun:14} and GOS \cite{Nedeljkovic:17} have exhibited minimum propagation losses below 3 dB/cm.

An alternative approach to changing the material is to use SOI wafers, but to limit the optical mode overlap with SiO$_2$, such as by suspending the Si waveguide core or by substantially increasing the Si core size \cite{Malik:13}. Compared to other platforms SOI wafers are easily available and the fabrication techniques are very well known, and as such at MIR wavelengths where Si is still transparent it may be preferred. In \cite{Xia:13Membrane,Cheng:12Membrane,Xiao:17} two dry-etch steps were used to first define a Si (or Ge) rib waveguide, and then to create an array of holes away from the waveguide mode that enable the buried oxide (BOX) to be locally removed by wet etching in hydrofluoric acid (HF). In \cite{Penades:16,SolerPenades:14}, we presented an alternative approach, consisting of a strip waveguide with subwavelength grating (SWG) holes that provide lateral optical confinement as well as access to the BOX. The concept of subwavelength grating (SWG) engineering of Si-wire waveguides was first proposed in \cite{Cheben:06, Cheben:10} and, more recently, extensively used in a variety of photonic devices \cite{Halir:15}. With this approach we achieved a minimum loss of 0.82 dB/cm at 3.8 $\mu$m. This design allows simpler fabrication consisting of a single dry etch step and it has the further advantage of reducing the necessary suspended region width and improving the mechanical stability, allowing the definition of wider devices like tapers and MMIs \cite{Penades:16}. The same approach has recently been implemented to create suspended Si slot waveguides \cite{Zhou:17} at 2.25 $\mu$m. 

We note that a further advantage of this waveguide platform compared to GOS and graded Si/Ge is the symmetrical high refractive index contrast between the core and upper and lower claddings. For example, GOS exhibits high index contrast between the Ge core (n = 4.0) and air upper cladding (n = 1.0), but low contrast with the Si lower cladding (n = 3.4), which results in the need for single mode waveguides to be rib waveguides with thickness on the order of several micrometers, and also results in very little mode overlap with the air upper cladding. Suspended Si waveguides would have a thinner waveguide core and larger evanescent field, and therefore much increased sensitivity for sensing applications where high mode overlap with an analyte placed on the waveguide surface is desirable.

To our knowledge, suspended Si waveguides have not yet been demonstrated at wavelengths beyond 3.8 $\mu$m. In this paper we report suspended silicon waveguides with 3.1 dB/cm propagation losses, 90$^o$ bends and s-bends at  $\lambda$ = 7.67 $\mu$m. At this wavelength bulk Si has a material absorption of $\sim$2.1 dB/cm \cite{ChandlerHorowitz:05}, increasing to 4.6 dB/cm at 9 $\mu$m and 9.3 dB/cm at 11 $\mu$m, thus the demonstrated waveguides in this letter approach the transparency limit of Si.

\begin{figure}[htbp]
	\centering
	\includegraphics[width=\linewidth]{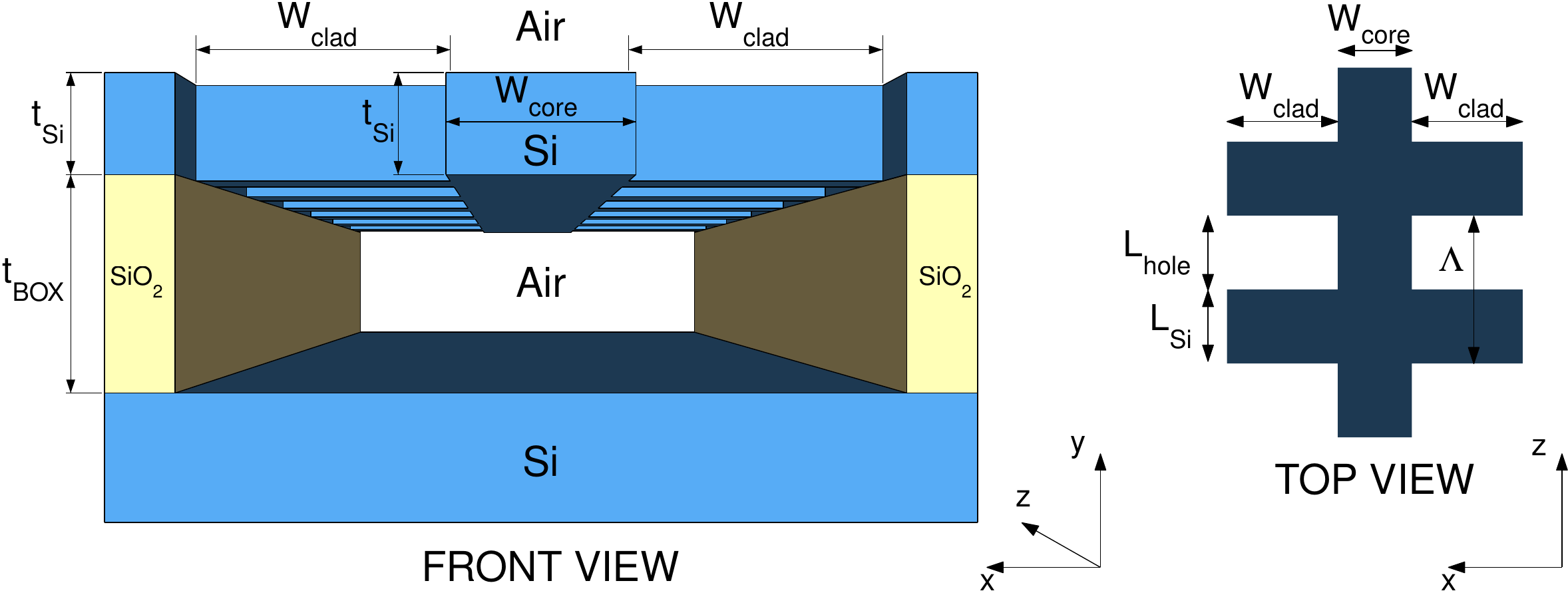}
	\caption{Schematic representation (front and top views) of the proposed suspended Si waveguide with subwavelength grating cladding.}
	\label{fig:SubWavWGScheme}
\end{figure}

A schematic representation of the proposed suspended waveguide is shown in Fig. \ref{fig:SubWavWGScheme}. This guiding structure, based on a SOI platform, comprises a solid silicon core of thickness $t_{Si}$ and width $W_{core}$ with a SWG structure of width $W_{clad}$ acting as the waveguide cladding \cite{Halir:09}. This lattice with holes and silicon strips of lengths $L_{hole}$ and $L_{Si}$, respectively, behaves as a homogeneous medium with an equivalent refractive index $n_{SWG}$, offering the index contrast required for waveguiding. In addition, the cladding is used to anchor the waveguide core to the unsuspended lateral silicon areas and to allow the entrance of the HF acid needed to remove the $t_{BOX}$-thick buried oxide layer. 

In our previous design working at $\lambda$ = 3.8 $\mu$m \cite{Penades:16}, we used a SOI wafer with $t_{Si}$ = 500 nm and $t_{BOX}$ = 3 $\mu$m. The waveguide core width was $W_{core}$ = 1.3 $\mu$m. The SWG cladding parameters ($W_{clad}$ = 2.5 $\mu$m, $L_{Si}$ = 100 nm and $L_{hole}$ = 450 nm) were optimized to provide mechanical stability to the entire structure and, simultaneously, facilitate the HF flow through the lattice holes. Scaling all dimensions by a factor of 2 could be acceptable to move from $\lambda$ = 3.8 $\mu$m to $\lambda$ = 7.67 $\mu$m, but this would imply $t_{BOX}$ = 6 $\mu$m. However, as in the long-wave range we use a SOI wafer with the same SiO$_2$ layer as before, $t_{BOX}$ thickness remains fixed to 3 $\mu$m. If all dimensions except $t_{BOX}$ are doubled, the power leakage to the Si substrate could lead to excessive propagation losses. Hence, the waveguide must be completely redesigned. 

We follow a design methodology based on two separate steps. Firstly, the silicon thickness $t_{Si}$ is chosen to have vertical single-mode operation with negligible leakage to the substrate. Secondly, both the core width $W_{core}$ and the SWG cladding ($W_{clad}$, $L_{Si}$ and $L_{hole}$) are adjusted to mitigate power leakage to the lateral Si without under-etching.

To study the vertical and lateral leakage losses of the TE-polarized modes supported by the waveguides, we used Rsoft™ and Photon Design\textsuperscript{\textregistered} simulation suites in combination with our 2D in-house electromagnetic tool FEXEN \cite{ZavargoPeche:12}, which is especially appropriate for designing periodic structures such as grating couplers and the SWG cladding of the waveguides.

\begin{figure}[htbp]
	\centering
	\includegraphics[width=\linewidth]{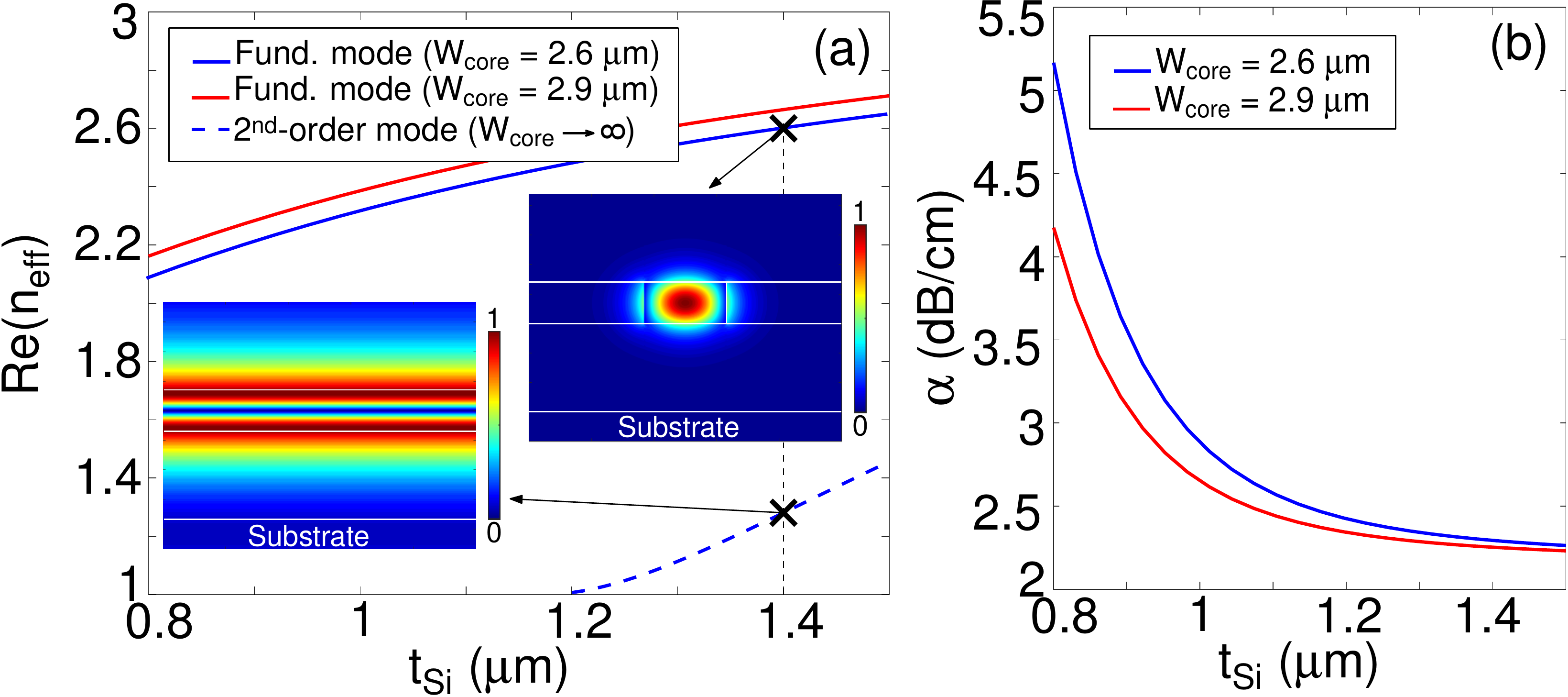}
	\caption{(a) Real part of the effective indices ($n_{eff}$) vs. the Si thickness ($t_{Si}$) for the fundamental mode ($W_{core}$ = 2.6 $\mu$m) and the second-order vertical mode ($W_{core} \rightarrow \infty$) supported by the suspended waveguides. Simulation data: $t_{BOX}$ = 3 $\mu$m, $n_{SWG}$ = 1.72. The real part of the effective index of the final designed waveguide with $W_{core}$ = 2.9 $\mu$m is also included. Insets: transverse field distributions for the first two modes ($W_{core}$ = 2.6 $\mu$m, $W_{core} \rightarrow \infty$) when $t_{Si}$ = 1.4 $\mu$m. (b) Power loss of the fundamental vertical mode of the suspended waveguide with $W_{core}$ = 2.6 $\mu$m and the finally designed waveguide as a function of the waveguide silicon thickness $t_{Si}$.}
	\label{fig:WGsims}
\end{figure}

Choosing the silicon thickness is critical to enable waveguiding with minimized propagation losses. A high value of $t_{Si}$ is beneficial to vertically confine the fundamental mode with low leakage to the substrate. On the other hand, a reduced silicon thickness is desirable to cut off the second-order vertical guided mode, or, at least, to ensure it leaks towards the substrate in the shortest possible propagation distance. In our initial simulations we kept the same SWG equivalent refractive index $n_{SWG}=1.72$ as in our original design \cite{Penades:16}. For the fundamental mode, we used a waveguide width $W_{core}$ = 2.6 $\mu$m, which coincides with the scaled value we employed at $\lambda$ = 3.8 $\mu$m. This value of 2.6 $\mu$m will be later optimized in the last step of the design process. Regarding the second-order vertical mode, it must be noticed that this mode should be avoided not only in interconnecting waveguides, but also in much wider optical devices like multimode interferometer (MMI) couplers. Since the vertical confinement for a wide waveguide is always higher than for a narrow waveguide, we have decided to calculate the effective index of the second-order vertical mode for a slab (i.e. $W_{core} \rightarrow \infty$) as a worst case. This provides us an upper bound, since any finite-width waveguide will exhibit a weaker guidance and it will therefore suffer higher substrate leakage losses. Figure \ref{fig:WGsims}(a) shows the variation of the real part of the effective indices of the fundamental and second-order vertical modes as a function of the waveguide core thickness $t_{Si}$. For $t_{Si}$ > $\sim$1 $\mu$m, the fundamental quasi-TE mode is well confined. As is observed, the second-order vertical mode is only weakly guided for Si thicknesses between $t_{Si}$ = 1.2 $\mu$m and $t_{Si}$ = 1.5 $\mu$m, as its low effective index, near 1 (refractive index of air), denotes the field is substantially expanded outside the waveguide silicon core. So, the second-order mode suffers from high vertical leakage in the depicted range. Figure \ref{fig:WGsims}(a) (inset) shows, for $t_{Si}$ = 1.4 $\mu$m, the transverse distributions of both the fundamental mode ($W_{core}$= 2.6 $\mu$m) and second-order vertical mode ($W_{core} \rightarrow \infty$).

To design an exact $t_{Si}$ value, a further examination of the propagation losses $\alpha$ is required. They have two contributions: the intrinsic material loss, which is $\alpha_{Si} \approx$ 2.1 dB/cm at $\lambda$ = 7.67 $\mu$m \cite{ChandlerHorowitz:05}, and the vertical power leakage to the substrate. In our simulations, we included silicon losses in the waveguide core and substrate. The total loss of the fundamental mode of a waveguide with $W_{core}$ = 2.6 $\mu$m is shown in Fig. \ref{fig:WGsims}(b) as a function of the core thickness $t_{Si}$. For Si thicknesses smaller than $t_{Si}$ $\sim$1 $\mu$m, substrate leakage losses are dominant and $\alpha$ exhibits an exponential increase. On the contrary, for larger thicknesses, propagation losses asymptotically tend to the material absorption $\alpha_{Si}$, as the mode is well confined for higher $t_{Si}$ values (> 1.3 $\mu$m). Furthermore, to increase the robustness of the design, the sensitivity of $\alpha$ with respect to $t_{Si}$ at the selected nominal Si thickness should be close to zero; otherwise, propagation losses could be rapidly increased if undesired reduction in the silicon thickness is produced by a HF overexposure during the fabrication process. Taking these considerations into account, we selected $t_{Si}$ = 1.4 $\mu$m. For this value, the fundamental mode leakage to substrate is negligible, more than ten times smaller than the intrinsic silicon losses. Although for the chosen $t_{Si}$ value the vertical second-order mode is indeed guided (see Fig. \ref{fig:WGsims}(a)), we can consider single-mode operation, as its propagation losses are greater than 50 dB/cm. Higher $t_{Si}$ values would barely benefit the fundamental mode loss, but undesirably decrease the second-order mode leakage. 

The waveguide core width and the SWG cladding geometry are chosen to avoid lateral leakage and to enhance the mechanical stability. Note that a scaled version of the original design in \cite{Penades:16}, i.e. $W_{core}$ = 2.6 $\mu$m and $W_{clad}$ = 5 $\mu$m, would compromise waveguide stability because of the thicker Si thickness of 1.4 $\mu$m. Therefore, we made the waveguide core wider ($W_{core}$ = 2.9 $\mu$m), increasing the lateral mode confinement and allowing a reduction in the cladding width up to $W_{clad}$ = 3 $\mu$m, with the intention of reducing the torque without degrading the lateral leakage. It is also important to mention that, even though all dimensions are substantially enlarged to migrate to $\lambda$ = 7.67 $\mu$m, the cladding width hardly increases with respect to the design at $\lambda$ = 3.8 $\mu$m, resulting in a remarkably more stable waveguide. At $\lambda$ = 7.67 $\mu$m, the Bragg period is moved away up to $\Lambda_{Bragg}\approx$ 1300 nm for a wide waveguide, thereby extending the subwavelength regime in relation to the design in \cite{Penades:16}. Due to the higher $t_{Si}$ value, larger holes are needed to assure an effective dry etching of the cladding and let HF penetrate through them. Therefore, we selected $L_{hole}$ = 900 nm. Finally, the fundamental mode lateral confinement and the waveguide mechanical stability were slightly improved by choosing $L_{Si}$ = 250 nm.

We simulated 90°-bends. Negligible losses are achieved for bend radius of 35 $\mu$m. S-bends were also designed, with propagation losses smaller than 0.04 dB/bend for an offset of 5 $\mu$m and a length of 75 $\mu$m.

To couple light from the optical fiber to the chip, we designed, for the first time, a focusing suspended surface grating coupler at a nominal wavelength of $\lambda$ = 7.67 $\mu$m. It is composed of a silicon membrane of thickness $t_{Si}$ where gaps are etched in the area where light should be coupled. The gaps again have a double purpose: they allow the removal of the silicon dioxide and constitute a periodic radiating waveguide formed by the succession of silicon and air wide strips. We achieve a simulated coupling efficiency of 58\% at the central wavelength and a 1-dB bandwidth of 230 nm. Very low back-reflections around 0.1\% are also achieved by including a thinner air gap at the beginning of the grating. The radiation angle is $\sim$19.3°. More details about the theoretical design process of this grating coupler and a full experimental characterization will be given in a future work.

Fabrication was carried out on a 6-inch SOI wafer. We implemented the same fabrication process as described in \cite{Penades:16}: e-beam lithography was used for patterning with ZEP-520A as positive resist, and etching was accomplished with an ICP tool using SF$_6$/C$_4$F$_8$ chemistry. The wafer was then submerged into a 1:7 HF bath for 2.5 hours, to ensure removal of the BOX beneath the widest devices. The resist was only removed after the HF etching in order to protect the Si top surface, as it was observed that this long wet etch produced a significant roughening of the Si surfaces. In our previous work \cite{Penades:16} it was observed that the dry and wet etches both resulted in some lateral etching, reducing the width of the silicon supports and the waveguide core thickness. To compensate for this a positive biasing of 220 nm (i.e. increase of the dimensions) was applied to the designed layout of all the unetched features. 

\begin{figure}[htbp]
	\centering
	\includegraphics[width=\linewidth]{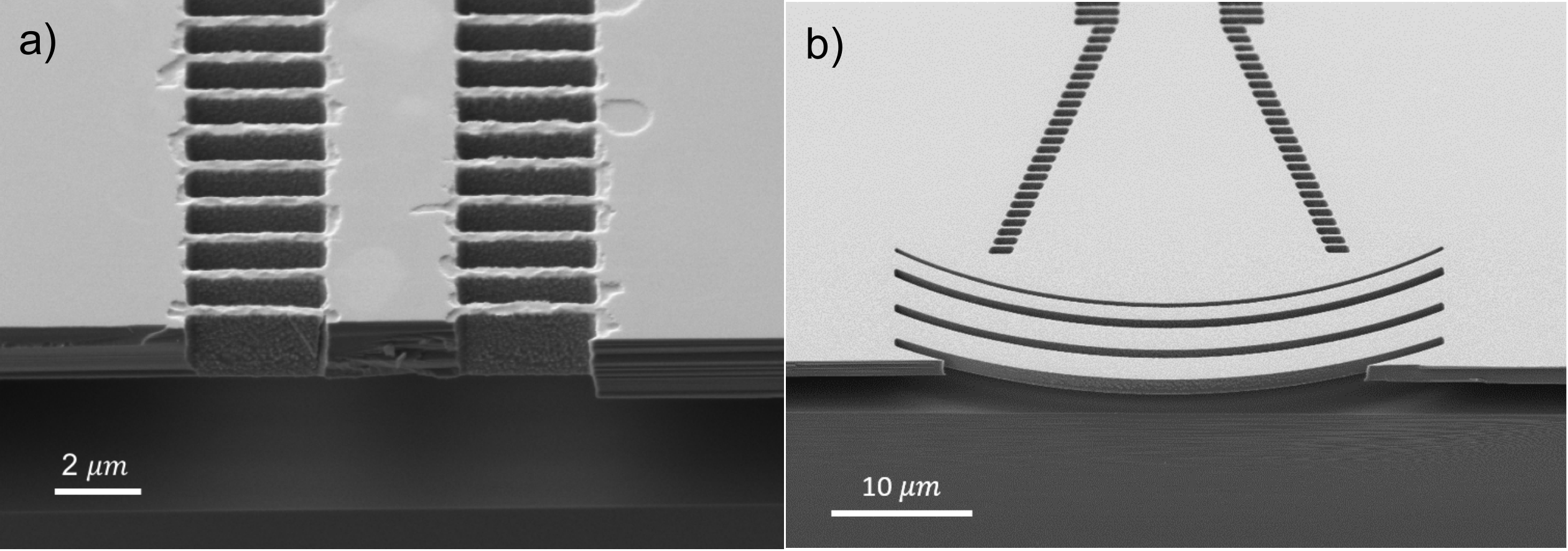}
	\caption{a) SEM image of a cleaved waveguide facet after fabrication. The BOX can be seen to have been completely removed below the waveguide, but the HF etching has created roughness along the etched edges. b) SEM image of the focusing grating coupler and taper with the BOX removed. It has been cleaved perpendicular to the waveguide, part way through the grating. The grating was the widest structure on the chip at 40 $\mu$m width.}
	\label{fig:SubWavSEMWG}
\end{figure}

Figure \ref{fig:SubWavSEMWG}(a) shows a scanning electron microscope (SEM) image of a fabricated waveguide, and Fig. \ref{fig:SubWavSEMWG}(b) shows a fabricated grating coupler. It can be seen that both devices are fully suspended and that there is no significant bending of the free standing Si. However, it can be seen that in some places roughness has been created along the edges of the etched structures in places where the HF was able to reach between the Si and photoresist layers during wet etching. Atomic force microscope measurements show that in these places 40 nm deep grooves have been etched into the Si. The final thickness of the Si waveguide, as measured using the SEM, was approximately 1420 nm.

The experimental setup employed a single mode continuous wave distributed feedback quantum cascade laser (Thorlabs QD7500CM1) with 106 mW maximum output power at a wavelength of 7.67 $\mu$m. A black diamond-2 lens with a focal length of 1.9 mm was used to collimate the light from the laser, a chopper wheel modulated the light and another black diamond-2 lens with a focal length of 6.0 mm was used to couple it into a single mode As$_2$Se$_3$ fiber (Coractive IRT-SE-28/170) with a measured output of $\sim$ 3.4 mW. Light was coupled into the waveguides with grating couplers (Fig. \ref{fig:SubWavSEMWG}(b)) and the output was collected via another single mode fiber and coupled to a liquid nitrogen cooled HgCdTe detector (Infrared Associates Inc. MCT-13-1.00). The signal from the detector was passed through a pre-amplifier before being sent to a lock-in amplifier to improve the signal to noise ratio. In order to characterize the dynamic range of the setup we placed a gold mirror in the position of a chip, finding that the maximum transmission was 60 dB above the noise floor.

\begin{figure}[htbp]
	\centering
	\includegraphics[width=\linewidth]{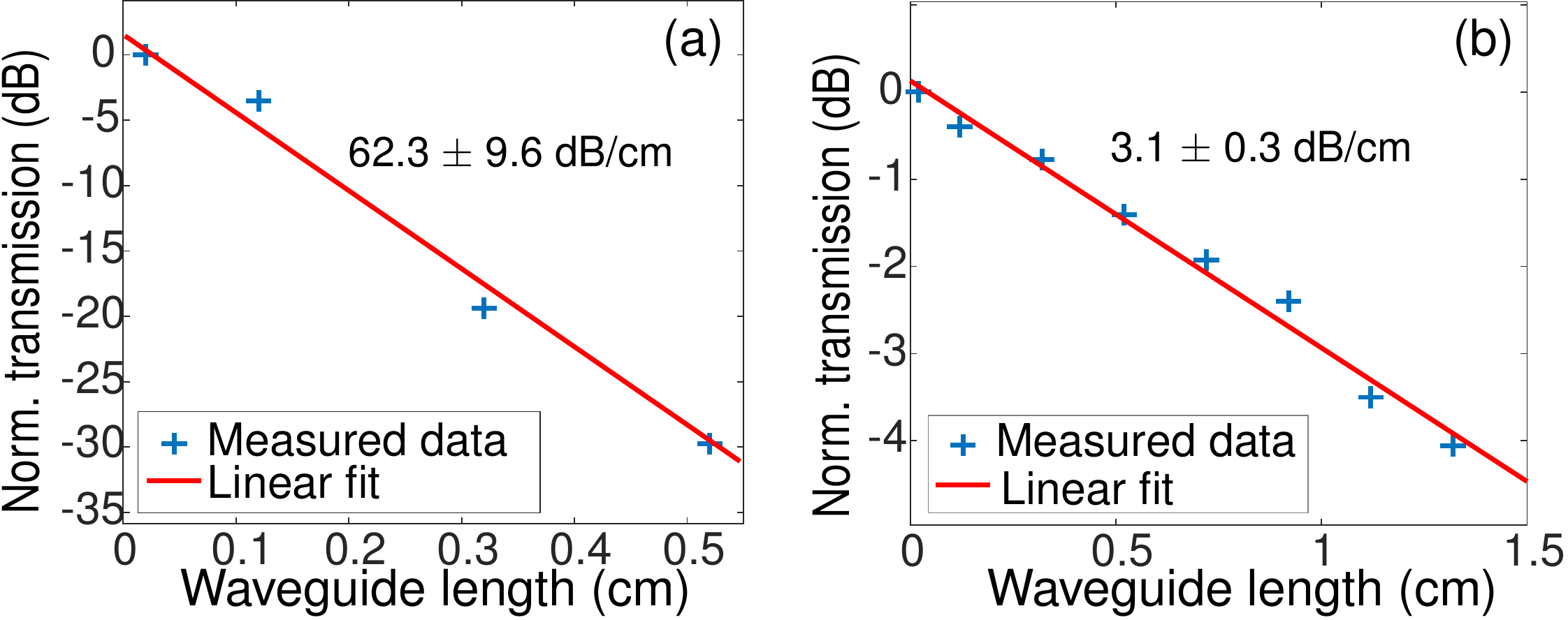}
	\caption{Waveguide cut-back loss measurement results at 7.67 $\mu$m wavelength for (a) Si waveguides with the BOX still present, and (b) suspended Si waveguides. For each set of points the transmissions have been normalized to the transmission through the shortest waveguide (0.02 cm) in each set. The red lines are linear fits to the measured points (cross symbols).}
	\label{fig:SubWavRes}
\end{figure}

The waveguide propagation loss was measured with the effective "cut-back" method. Figures \ref{fig:SubWavRes}(a) and \ref{fig:SubWavRes}(b) show the transmissions of waveguides of different lengths in samples before and after BOX removal, respectively. The figures show that removing the SiO$_2$ reduces the waveguide loss from $62.3\pm9.6$ dB/cm to $3.1\pm0.3$ dB/cm. According to simulations 2.2 dB/cm of the suspended waveguide loss can be attributed to material loss ($\sim$2.1 dB/cm) and substrate leakage ($\sim$0.1 dB/cm), therefore approximately 0.9 dB/cm of the loss is likely due to scattering, predominantly from the roughness created by HF etching. This waveguide loss is comparable to other group-IV material waveguide demonstrations in this wavelength range \cite{Brun:14, Nedeljkovic:17, Vakarin:17}, and we expect that it can be reduced further through optimization of the HF etching process. 

The 90° bend loss was 0.08$\pm$0.02 dB/bend and the loss of the s-bends was 0.06$\pm$0.02 dB/bend.

We have designed, fabricated and characterized suspended silicon waveguides, 90$^o$ bends and s-bends at a wavelength of 7.67 $\mu$m. The propagation loss for the waveguides was $3.1\pm0.3$ dB/cm while the bend loss was 0.08$\pm$0.02 dB/bend and 0.06$\pm$0.02 dB/bend for the 90$^o$ bends and s-bends respectively. The material loss of Si contributes by $\sim$2.1 dB/cm to the propagation loss figure, and it is expected that the remainder of the loss predominantly comes from scattering at the roughness that prolonged HF etching creates at the Si/air interfaces. Our future work will focus on refining the fabrication process in order to reduce roughness and to approach propagation loss values closer to the material loss. These results demonstrate that suspended silicon waveguides designed using the sub-wavelength grating concept and fabricated from SOI wafers can be used throughout the mid-infrared transparency window of silicon, and unlock potential sensing applications in this range.

\bigskip
\noindent
\textbf{Acknowledgements.} The data used to create Figs. \ref{fig:WGsims} and \ref{fig:SubWavRes} can be accessed online at http://doi.org/10.5258/SOTON/D0322

\bigskip
\noindent
\textbf{Funding.} Engineering and Physical Sciences Research Council (EPSRC) (MIGRATION EP/L01162X/1, National Hub in High Value Photonic Manufacturing EP/N00762X/1, CORNERSTONE EP/L021129/1); Royal Academy of Engineering (RAEng) (M. Nedeljkovic fellowship RF201617/16/33); Ministerio de Economía y Competitividad (MINECO) (TEC2016-80718-R); Ministerio de Educación, Cultura y Deporte (MECD) (FPU14/06121); Fondo Europeo de Desarrollo Regional – FEDER; Universidad de Málaga.

\bigskip


\bibliography{sample}

\bibliographyfullrefs{sample}
 

\end{document}